\documentclass[aps,prl,showpacs,twocolumn]{revtex4}
\usepackage{graphicx}
\usepackage{amsmath}
\usepackage{amsfonts}
\usepackage{color}
\usepackage[normalem]{ulem}

\begin{document}

\title{Edge Excitations in Fractional Chern Insulators}
\author{Wei-Wei Luo$^{1}$, Wen-Chao Chen$^{1}$, Yi-Fei Wang$^{1}$, Chang-De Gong$^{1,2}$ }
\affiliation{$^1$Center for Statistical and Theoretical Condensed Matter Physics, and Department of Physics, Zhejiang Normal University, Jinhua 321004, China \\$^2$National Laboratory of Solid State Microstructures and Department of Physics, Nanjing University, Nanjing 210093, China}
\date{\today}

\begin{abstract}
Recent theoretical works have demonstrated the realization of fractional quantum anomalous Hall states (also called fractional Chern insulators) in topological flat band lattice models without an external magnetic field. Such newly proposed lattice systems play a vital role to obtain a large class of fractional topological phases. Here we report the exact numerical studies of edge excitations for such systems in a disk geometry loaded with hard-core bosons, which will serve as a more viable experimental probe for such topologically ordered states. We find convincing numerical evidence of a series of edge excitations characterized by the chiral Luttinger liquid theory for the bosonic fractional Chern insulators in both the honeycomb disk Haldane model and the kagom\'{e}-lattice disk model. We further verify these current-carrying chiral edge states by inserting a central flux to test their compressibility.
\end{abstract}

  \maketitle

{\it Introduction.---} One of the most essential and fascinating topics in condensed matter physics is to explore and classify the various states of matter, among which the integer quantum Hall effect (IQHE) \cite{Klitzing} and the fractional quantum Hall effect (FQHE) \cite{Tsui} have long been the major focus. It is well known that, at fractional fillings, the FQHE will emerge when interacting particles move in Landau levels (LLs) caused by an external uniform magnetic field. On the other hand, great interest has been aroused to realize both the IQHE and FQHE in lattice models in the absence of an external magnetic field. An initial theoretical attempt was made by Haldane \cite{Haldane}, who proposed a prototype lattice model to achieve the IQHE by introducing two non-trivial topological bands with Chern numbers $C=\pm 1$~\cite{Thouless}. Haldane's model demonstrates that the IQHE can also be attained without LLs, i.e. defines the quantum anomalous Hall (QAH) states. The lattice version of the FQHE without LLs, however, comes much latter because of its intriguing strongly correlated nature.

The recent proposal of topological flat bands (TFBs) fulfils the basic ingredients to explore the intriguing fractionalization phenomenon without LLs. TFBs~\cite{TFBs} belong to a class of band structures with at least one nearly flat band with non-zero Chern number. This may be viewed as the lattice counterpart of the continuum LLs. Several systematic numerical works have been done to explore the correlation phenomenon within TFB models, and Abelian \cite{Sheng1,YFWang1,Regnault2} and non-Abelian \cite{YFWang2,Bernevig,Bernevig2} FQHE states without LLs have already been well established numerically.
This intriguing fractionalization effect in TFBs without an external magnetic field, defines a new class of fractional topological phases, i.e. fractional quantum anomalous Hall (FQAH) states, now also known as fractional Chern insulators (FCIs).
Some new approaches, e.g. the Wannier-basis model wave functions and pseudo-potentials~\cite{XLQi}, the projected density operator algebra~\cite{Sondhi}, the parton wave-function constructions \cite{parton}, and also the adiabatic continuity paths~\cite{continuity}, have been proposed to further understand these FCI/FQAH states in TFBs.
There are various other proposals of TFB models and material realization schemes \cite{Fiete,Xiao,Venderbos,Ghaemi,FWang,Venderbos2,Weeks,RLiu,WCChen,Bergholtz,SYang2,Grushin,
Lukin,Yannopapas,CWZhang,FLiu,Cirac,Cooper}.
Very recent systematic numerical studies have also found exotic FCI/FQAH states in TFBs with higher Chern numbers ($C\ge2$) which do not have the direct continuum analogy in LLs~\cite{YFWang3,ZLiu,Sterdyniak}, and also the hierarchy FCI/FQAH states~\cite{Hierarchy}.

\begin{figure}[!tb]
  \begin{minipage}[c]{0.5\textwidth}
  \includegraphics[scale=0.5]{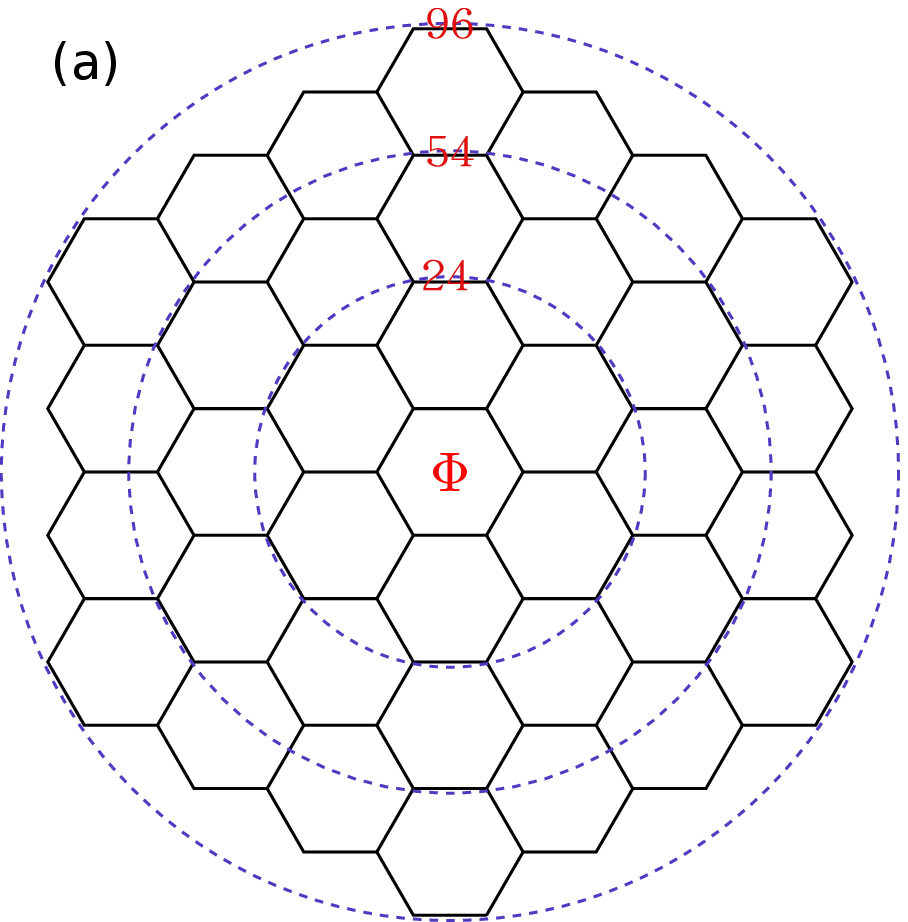}
\hspace{5in}
\end{minipage}
  \begin{minipage}[c]{0.5\textwidth}
  \includegraphics[scale=0.6]{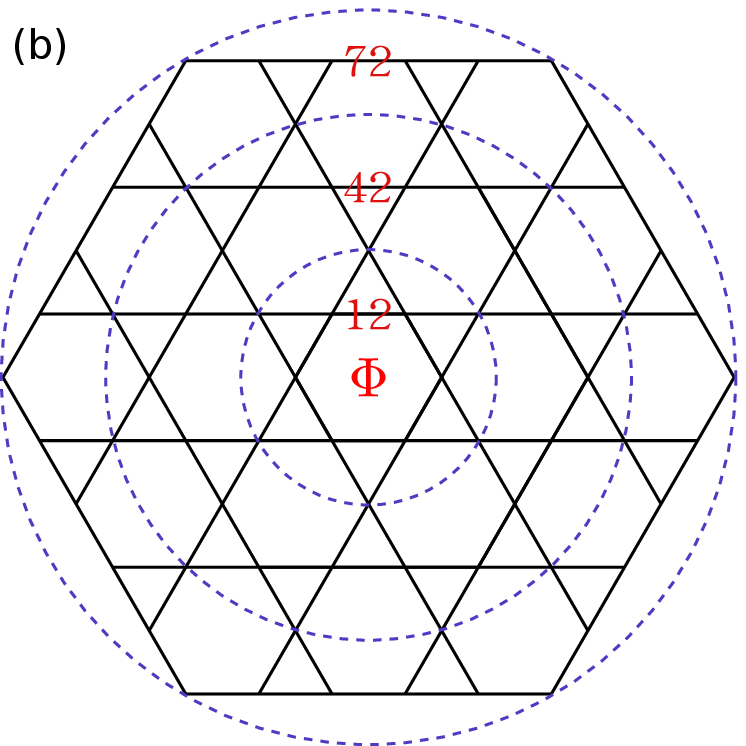}
\end{minipage}
  \caption{(color online). (a) The honeycomb-lattice disk and (b) the kagom\'{e}-lattice disk. Both disk geometries satisfy the $C_6$ rotational symmetry. Various disk sizes are indicated by the circles and the labelled numbers.}
\label{disk}
\end{figure}

Although previous studies have firmly established various properties of a large class of FCI/FQAH phases, we are here concerned with the less studied edge excitations since they in principle should open another window to reveal the bulk topological order~\cite{XGWen}. Edge excitations might also provide a more viable experimental probe especially considering about the possible future realizations of FCI/FQAH phases in optical lattice systems~\cite{Lukin,Cooper}. Some very promising and detailed experimental schemes have been proposed for realizing the TFB models and the FCI/FQAH phases, e.g. implementing dipolar spin systems with ultracold polar molecules trapped in a deep optical lattice driven by spatially modulated electromagnetic fields~\cite{Lukin}. The chiral edge modes of such topological phases are expected to be directly visualized in optical-lattice-based experiments~\cite{Stanescu}. A recent work has studied edge excitations of the bosonic FQHE in LLs generated by an artificial uniform flux in optical lattices~\cite{Kjall}, where finite-size disks and cylinders with trap potentials have been considered, and the edge spectrum is clearly observed. Also, related knowledge of FQHE edge states also exists in the study of orbital entanglement spectra~\cite{entanglement}. In the present work, we investigate edge excitations of the bosonic FCI/FQAH phases~\cite{YFWang1} of TFB models in a disk geometry. Through extensive systematic numerical exact diagonalization (ED) studies, we demonstrate clear evidence of edge excitation spectra of bosonic FCI/FQAH phases. These edge excitation spectra are quite coincident with the chiral Luttinger liquid theory \cite{XGWen}. To further check the compressibility of these edge states, we insert a central flux into the disk system and indeed verify these current-carrying chiral edge states upon tuning the flux strength.

{\it Formulation.---} We first look into the Haldane model~\cite{Haldane} on a honeycomb-lattice disk, which is loaded with interacting hard-core bosons~\cite{YFWang1}:
\begin{eqnarray}
H_{\rm HC}= &-&t^{\prime}\sum_{\langle\langle\mathbf{r}\mathbf{r}^{
\prime}\rangle\rangle}
\left[b^{\dagger}_{\mathbf{r}^{ \prime}}b_{\mathbf{r}}\exp\left(i\phi_{\mathbf{r}^{ \prime}\mathbf{r}}\right)+\textrm{H.c.}\right]\\
&-&t\sum_{\langle\mathbf{r}\mathbf{r}^{ \prime}\rangle}
\left[b^{\dagger}_{\mathbf{r}^{\prime}}b_{\mathbf{r}}+\textrm{H.c.}\right]
-t^{\prime\prime}\sum_{\langle\langle\langle\mathbf{r}\mathbf{r}^{
\prime}\rangle\rangle\rangle}
\left[b^{\dagger}_{\mathbf{r}^{\prime}}b_{\mathbf{r}}+\textrm{H.c.}\right]\nonumber
\label{e.1}
\end{eqnarray}
where $b^{\dagger}_{\mathbf{r}}$ creates a hard-core boson at site $\mathbf{r}$,  $\langle\dots\rangle$, $\langle\langle\dots\rangle\rangle$ and $\langle\langle\langle\dots\rangle\rangle\rangle$ denote the NN, the NNN and the next-next-nearest-neighbor (NNNN) pairs of sites, respectively. We adopt the previous parameters~\cite{YFWang1} to achieve a lowest TFB with a flatness ratio of about 50 (the ratio of the band gap over bandwidth): $t=1$, $t^{\prime}=0.60$, $t^{\prime\prime}=-0.58$ and $\phi=0.4\pi$.

We also consider a kagom\'{e} lattice TFB model~\cite{TFBs,RLiu} which is also loaded with hard-core bosons, and the model Hamiltonian has the form:
\begin{eqnarray}
H_{\rm KG}= &-&t\sum_{\langle\mathbf{r}\mathbf{r}^{ \prime}\rangle}
\left[b^{\dagger}_{\mathbf{r}^{ \prime}}b_{\mathbf{r}}\exp\left(i\phi_{\mathbf{r}^{ \prime}\mathbf{r}}\right)+\textrm{H.c.}\right]\nonumber\\
&-&t^{\prime}\sum_{\langle\langle\mathbf{r}\mathbf{r}^{\prime}\rangle\rangle}
\left[b^{\dagger}_{\mathbf{r}^{\prime}}b_{\mathbf{r}}+\textrm{H.c.}\right]
\label{e.2}
\end{eqnarray}
We choose the previous parameters for this kagom\'{e}-lattice model~\cite{RLiu}: $t=1$, $t^{\prime}=-0.19$, $\phi=0.22\pi$, which leads to a lowest TFB with the flatness ratio of about 20.

Convincing numerical evidence of the $1/2$ bosonic FCI/FQAH states at the $1/2$ hard-core boson filling of a lowest TFB on a torus geometry has been well established previously~\cite{YFWang1}, with the formation of a quasi-degenerate ground-state (GS) manifold, characteristic GS momenta, a robust bulk excitation spectrum gap, as well as a fractional quantized Chern number for each GS. The disk geometries for both lattice models are illustrated in Fig. \ref{disk}, which shows the $C_6$ rotational symmetry. An additional trap potential is required on the finite-size disk systems to confine the FCI/FQAH droplet, outside which the edge modes are able to propagate around the disk. Here we choose the conventional harmonic trap, of the form $V = V_{\rm trap}\sum_{\mathbf{r}} |\mathbf{r}|^2 n_{\mathbf{r}}$ with $V_{\rm trap}$ as the potential strength~\cite{Kjall} (with the NN hopping $t$ as the energy unit), $|{\mathbf{r}}|$ as the radius from the disk center (with the half lattice constant $a/2$ as the length unit), and $n_{\mathbf{r}}$ as the boson number operator.

\begin{figure}[!htb]
  \includegraphics[scale=0.75]{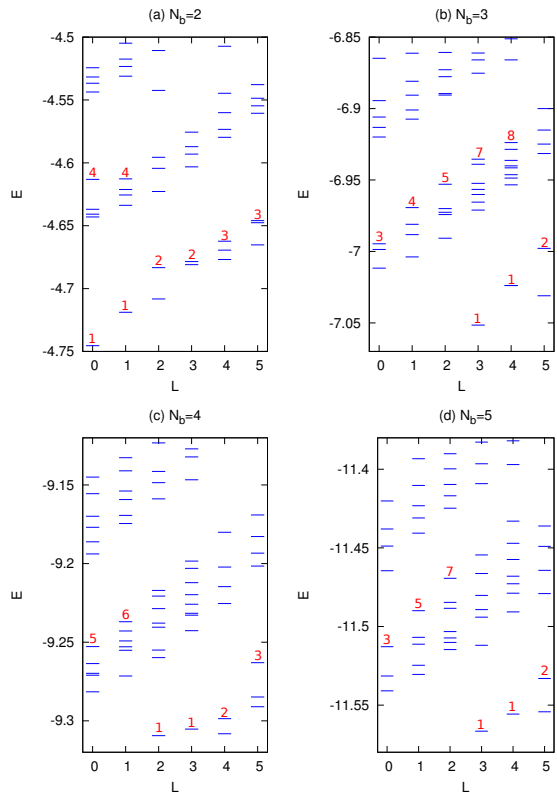}
  \caption{(color online). Edge excitations on a honeycomb disk with $N_s=96$ sites and trap potential $V_{\rm trap}=0.005$. Numbers labelled upon low energy levels in each sector show the quasi-degeneracy of low edge excitations, which is also the sequence derived from the chiral Luttinger liquid theory.}
\label{HC_edge}
\end{figure}

{\it Edge excitations.---}We have first considered the honeycomb disk model, and have observed clear edge excitation spectra for a relative broad range of the trap potential strength $V_{\rm trap}$ and various honeycomb disk sizes with $N_s = 24, 54, 96$ sites. Some representative results are shown in Fig. \ref{HC_edge}. As the system owns the $C_6$ rotational symmetry, each energy states can be classified with a quantum number of angular momentum $L = 0, 1, 2, 3,..., \mod 6$. These edge-excitation quasi-degeneracies of a finite lattice system with various boson numbers are expected to match with that predicted by the chiral Luttinger liquid theory for the lowest few sectors~\cite{Kjall}.

\begin{table}[h]
\begin{tabular}{c|l|l c c c c c}
 \hline
 \hline
$\Delta L$ & \{$n_1,n_2,n_3,n_4,n_5,n_6$,...\}        &  &2b &3b &4b &5b &6b  \\
 \hline
 \hline
     0 & \{0,0,0,0,0,0,...\}                          &$d=$  &1 &1 &1 &1  &1  \\

 \hline
     1 & \{1,0,0,0,0,0,...\}                          &$d=$  &1 &1 &1 &1  &1  \\
 \hline
     2 & \{0,1,0,0,0,0,...\},   \{2,0,0,0,0,0,...\}  &$d=$  &2 &2 &2 &2  &2  \\
 \hline
     3 & \{0,0,1,0,0,0,...\},   \{1,1,0,0,0,0,...\}  &$d=$  &2 &3 &3 &3  &3  \\
       & \{3,0,0,0,0,0,...\}                                                 \\
 \hline
     4 & \{0,0,0,1,0,0,...\},   \{1,0,1,0,0,0,...\}  &$d=$  &3 &4 &5 &5  &5  \\
       & \{0,2,0,0,0,0,...\},   \{2,1,0,0,0,0,...\}                         \\
       & \{4,0,0,0,0,0,...\}                                                 \\
 \hline
     5 & \{0,0,0,0,1,0,...\},   \{1,0,0,1,0,0,...\}  &$d=$  &3 &5 &6 &7  &7  \\
       & \{0,1,1,0,0,0,...\},   \{2,0,1,0,0,0,...\}                         \\
       & \{1,2,0,0,0,0,...\},   \{3,1,0,0,0,0,...\}                         \\
       & \{5,0,0,0,0,0,...\},                                                \\
 \hline
     6 & \{0,0,0,0,0,1,...\},   \{1,0,0,0,1,0,...\}  &$d=$  &4 &7 &\underline{9} &\underline{10} &\underline{11} \\
       & \{0,1,0,1,0,0,...\},   \{2,0,0,1,0,0,...\}                         \\
       & \{0,0,2,0,0,0,...\},   \{1,1,1,0,0,0,...\}                         \\
       & \{3,0,1,0,0,0,...\},   \{0,3,0,0,0,0,...\}                         \\
       & \{2,2,0,0,0,0,...\},   \{4,1,0,0,0,0,...\}                         \\
       & \{6,0,0,0,0,0,...\}                                                 \\
 \hline
 \hline
\end{tabular}
\caption{Occupation of chiral edge bosons in $k=1, 2, 3, 4, 5, 6,$ ... orbitals: \{$n_1,n_2,n_3,n_4,n_5,n_6$, ...\}. $d$ is the degeneracy of the edge excitations in a sector. $\Delta L = L - L_{\rm GS} = \sum_{k} kn_{k}$, which is the shifted total angular momentum relative to the ground state angular momentum. The right column shows the degeneracy in systems with finite boson numbers, e.g., for the ``4b'' column ($N_b=4$), the predicted degeneracy is  $1, 1, 2, 3, 5, 6, 9, ...$ in $\Delta L=0,1,2,3,4,5,6,...$ angular momentum sectors. The partial degeneracy sequences are mostly observed (except the underlined $\underline{9}$, $\underline{10}$ and $\underline{11}$) in our ED results.}
\label{luttinger}
\end{table}

According to the hydrodynamical approach~\cite{XGWen}, low energy edge excitations of a Laughlin-state FQHE droplet at filling $\nu=1/m$ are generated by the Kac-Moody algebra, and form a chiral Luttinger liquid with the effective Hamiltonian:
\begin{equation}\label{EdgeH}
   H_{1/m}=2\pi \frac{v}{\nu}\displaystyle\sum_{k>0}\rho_{-k}\rho_k=v\sum_{k>0} k a^{\dagger}_k a_k,
\end{equation}
with the angular momentum $k$ along the edge and the Fourier-transformed one-dimensional density operator $\rho_k$, $[\rho_k,\rho_{k'}]=\frac{\nu}{2\pi}k\delta_{k+k'}$, $a^{\dagger}_k$ and $a_k$ are chiral boson/phonon operators. Based on this theory, chiral edge bosons occupy in $k=1, 2, 3, 4, 5, 6, ...$ (unit $2\pi/M$ and edge length $M$) angular momentum orbitals with the occupation numbers \{$n_1,n_2,n_3,n_4,n_5,n_6$, ...\}. We denote $\Delta L = L - L_{\rm GS}$ as the shifted total angular momentum relative to the GS angular momentum. For a given $\Delta L = \sum_{k} kn_{k}$, it is easy to count the edge-state degeneracies by numerating the allowed occupation configurations. In the thermodynamics limit with infinite boson numbers, the degeneracy sequence should be 1,1,2,3,5,7,11,15,22,...

For our systems with only small boson number $N_b$'s, edge excitations made up of $\sum_k n_k\le N_b$ single modes are expected. For three bosons ($N_b=3$), after discarding some configurations in the middle column of Table \ref{luttinger} with $\sum_k n_k>N_b=3$, the predicted degeneracy should be 1,1,2,3,4,5,7,..., which is in good accordance with the observed quasi-degeneracy in our ED results [Fig. \ref{HC_edge}(b)]. For four bosons ($N_b=4$), after discarding some configurations with $\sum_k n_k>N_b=4$, the predicted degeneracy should be 1,1,2,3,5,6,..., which is also in good accordance with the observed quasi-degeneracy in our ED results [Fig. \ref{HC_edge}(c)]. The right column in Table \ref{luttinger} lists the partial degeneracy sequences which are mostly observed in our ED results.

\begin{figure}[!htb]
  \includegraphics[scale=0.75]{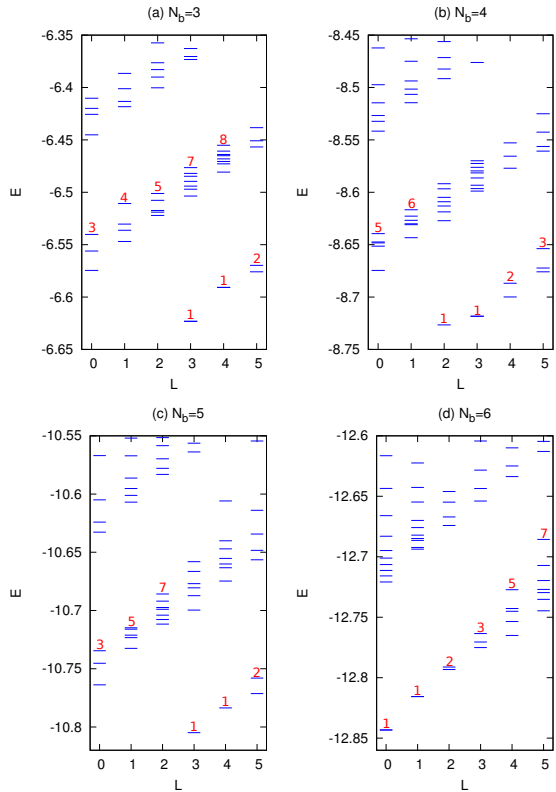}
  \caption{(color online). Edge excitations on a kagom\'{e} disk with $N_s=72$ sites and trap potential $V_{\rm trap}=0.005$. Numbers labelled upon low energy levels in each sector show the quasi-degeneracy of low edge excitations, which is also the sequence derived from the chiral Luttinger liquid theory.}
\label{KG_edge}
\end{figure}

Now we turn to investigate the kagom\'{e} disk model. We studied the finite system also in several different sizes, with $N_s =12, 30, 42, 72$ sites. Edge spectra from a finite kagom\'{e} disk with $N_s=72$ sites are shown in Fig. \ref{KG_edge}, which shows even clearer edge excitation spectra than that obtained from $N_s=96$ honeycomb disk. It is quite exciting to observe that these edge excitations are independent of lattice geometry even in such small systems. For an example with six bosons ($N_b=6$), after discarding some states with $\sum_k n_k > N_b=6$, the predicted degeneracy should be 1,1,2,3,5,7,... which is also in exact accordance with the observed quasi-degeneracy in our ED results  [Fig. \ref{KG_edge}(d)]. It is also observed that with the larger disk sizes and smaller boson numbers, the number of matched momentum sectors increases, and thus displays less severe finite size effects.

For both disk models, we have noticed that the GSs locate at various angular momenta for different boson numbers. The GS angular momenta can be heuristically analyzed from the generalized Pauli principle~\cite{Pauli,Regnault2}, which states that no more than one particle occupy two consecutive orbitals for $1/2$ FCI/FQAH phases, and proves to be a simple yet helpful tool in understanding numerical results~\cite{supply}.

\begin{figure}[!htb]
  \includegraphics[scale=0.5]{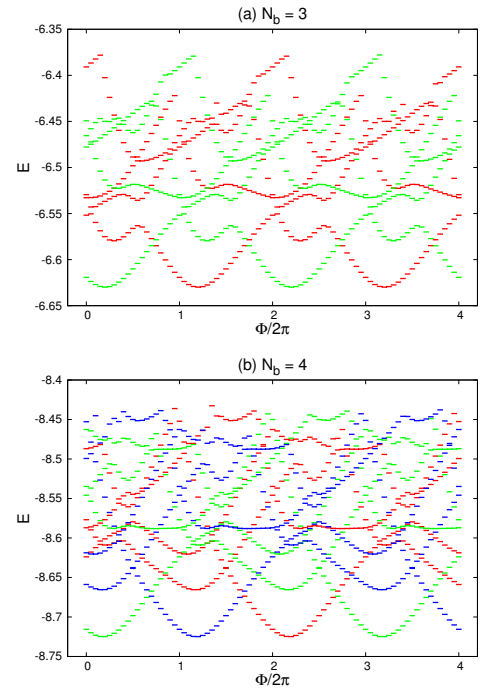}
  \caption{(color online). Evolution of low energy states with the varying central flux of the  kagom\'{e} disk with $N_s=42$ sites, trap potential $V_{\rm trap}=0.005$, and various boson numbers $N_b$'s.
  (a) For $N_b = 3$, low energy states in $L= 0, 3$ sectors (red, green respectively) evolve into each other.
  (b) For $N_b = 4$, low energy states in $L= 0, 2, 4$ sectors (red, green, blue respectively) evolve into each other.}
\label{flux}
\end{figure}

{\it Central flux.---}As a next step, we insert a flux into the center of disk geometry (as shown in Fig.~\ref{disk}) and tune the flux strength to test the compressibility of edge excitations. Consider the kagom\'{e} disk with $N_s=42$ sites for an example; the trap potential will always be fixed as $V_{\rm trap}=0.005$. For the system with $N_b=3$ bosons, low energy levels $E_n$ in $L= 0, 3$ sectors (or $L= 1, 4$ and $L= 2, 5$ sectors) evolve into each other as shown in Fig. \ref{flux}(a):
the low energy levels in the $L= 0$ sector (red) and those in the $L= 3$ sector (green) are
interchanged when the central flux $\Phi$ changes its value by $2\pi$, i.e. $E_n(L+3,\Phi+2\pi)=E_n(L,\Phi)$ (where $L,L+3=0,1,2,\dots\mod 6$). These energy spectra $E_n$ finally return to themselves after two periods of the central flux, i.e. $E_n(L,\Phi+4\pi)=E_n(L,\Phi)$.
It is obvious to observe the current-carrying nature, i.e. the compressibility of edge states since all of them evolve into the higher energy spectra without any gap.

For the system with $N_b=4$ bosons, low energy levels $E_n$ in $L= 0, 2, 4$ sectors (or $L= 1, 3, 5$ sectors) evolve into each other as is shown in Fig. \ref{flux}(b):
the low energy levels in the $L= 0$ sector (red), the $L= 2$ sector (green)
and the $L= 4$ sector (blue) are cyclically interchanged when the central flux $\Phi$ changes its value by $2\pi$ or $4\pi$, i.e. $E_n(L+2,\Phi+2\pi)=E_n(L+4,\Phi+4\pi)=E_n(L,\Phi)$ (where $L,L+2,L+4=0,1,2,\dots\mod 6$). These states finally return to themselves after three periods of the central flux, i.e. $E_n(L,\Phi+6\pi)=E_n(L,\Phi)$.
For our disk systems with the $C_6$ rotational symmetry, when boson numbers satisfy $N_b/6=p/q$ with $p$ and $q$ as coprime integers, there is a generic periodicity of energy spectra: $E_n(L,\Phi+2q\pi)=E_n(L,\Phi)$. Such periodicity can be heuristically understood by considering that the many-body states are built from single-particle orbitals with particular flux periodicity~\cite{supply}.

{\it Summary and discussion.---}We considered two representative TFB lattice models in disk geometry to explore edge excitations of FCI/FQAH states. With a confining harmonic trap potential, very clear edge excitation spectra are observed with the quasi-degeneracy counting rule satisfying the chiral Luttinger liquid theory at least for the lowest six sectors. By inserting and tuning a central flux to both systems, we are able to examine the compressibility of these current-carrying chiral edge states. Our work here focuses on the bosonic FCI/FQAH states in TFBs with the Chern number $C = 1$, while it is very interesting to explore the exotic edge excitations in high-Chern-number FCI/FQAH phases~\cite{YFWang3,ZLiu,Sterdyniak} and also those in the hierarchy FCI/FQAH phases~\cite{Hierarchy} in future works. We also expect that edge excitation spectra provide another window to reveal the bulk topological order, as well as a more viable experimental probe to future FCI/FQAH experimental systems.

We acknowledge Prof. D. N. Sheng for helpful discussions and previous collaborations.
This work is supported by the NSFC of China Grants No. 10904130, No. 11374265 (Y.F.W.)
and No. 11274276 (C.D.G.), and the State Key Program for Basic Researches of China
Grant No. 2009CB929504 (C.D.G.).

{\it Note added.---}After the completion of the present work, we became aware of a very recent related preprint~\cite{ZLiu2} which reports the orbital entanglement spectrum study of the bulk-edge correspondence and edge states in FCIs.

\newpage

\section*{SUPPLEMENTARY MATERIAL FOR ``EDGE EXCITATIONS IN FRACTIONAL CHERN INSULATORS''}

In the main text of this paper, we have obtained edge-state spectra
for finite-size disk systems with different boson numbers, whose ground
states locate at various angular momenta. Also the periodicity of the
energy spectra under flux insertion depends on particle numbers, which is
$E_n(L,\Phi+2q\pi)=E_n(L,\Phi)$ when boson numbers satisfy $N_b/6=p/q$
with $p$ and $q$ as coprime integers. We can indeed heuristically
understand these numerical results in view of the generalized Pauli
principle which rules the boson occupancy in single-particle orbitals with
particular angular momenta, and considering that many-body states are built
from single-particle orbitals with particular flux periodicity.

\begin{figure}[!htb]
    \includegraphics[scale=0.35]{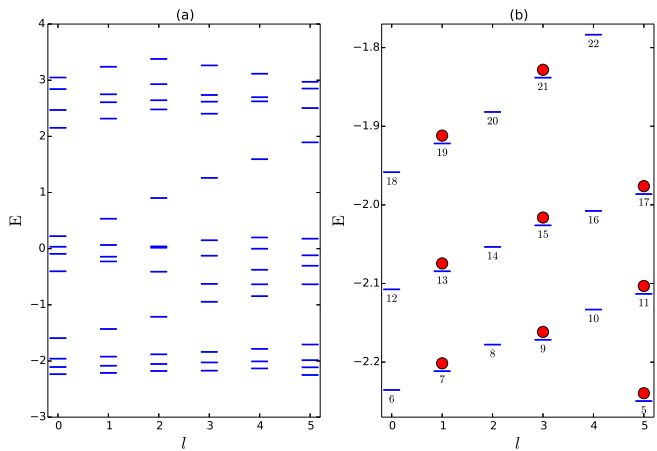}
    \caption{(color online). Single-particle orbitals on a
        kagom\'{e} disk with $N_s=72$ sites and the trap potential $V_{\rm
        trap}=0.005$.  (a) Single-particle energy spectrum.  (b) Zoom
        in on the lowest energy levels of (a). Red circles
        represents the hardcore bosons occupied in
        single-particle orbitals (the number labelled below each low
        energy level is the angular momentum quantum number), the
        configuration is ruled by the generalized Pauli
        principle. The ground state locates at the angular momentum $l =
        5$, with subsequent low-energy excited states lying in
incremental momentum sector $l = 4 + i$ where $i$ denotes $i$th lowest
energy state. }
\label{KG_1b_edge}
\end{figure}

\section*{Single-particle orbitals}

\begin{figure}[!htb]
    \includegraphics[scale=0.4]{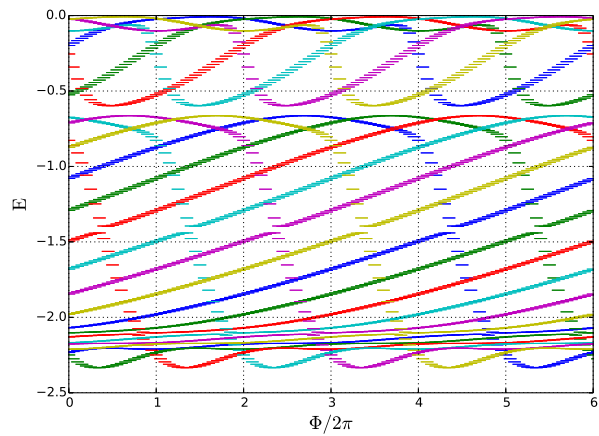}
    \caption{(color online). Evolution of single-particle low-energy
        states with the varying central flux of the  kagom\'{e} disk
        with $N_s=42$ sites and the trap potential $V_{\rm trap}=0.005$.
    A single-particle state with the angular momentum $l$ evolves into the
state with the angular momentum $l+1 \mod 6$. Different colors represents the energy
states from different angular momentum sectors.}

\label{1b_flux}
\end{figure}

Let us first look at the single-particle orbitals of these finite disk
systems, where we take kagom\'{e} lattice with 72 sites (corresponding
to the system studied in Fig. 3 of the main text) as an example.
Parameters for the kagom\'{e} lattice Hamiltonian are kept the same as
in the main text. The energy spectrum for this tiny system is
illustrated in Fig.~\ref{KG_1b_edge}, showing the single-particle ground state locating
at angular momentum $l = 5$, with subsequent low-energy excited states lying in
incremental momentum sector $l = 4 + i$ where $i$ denotes $i$th lowest
energy state. Considering $C_6$ rotational symmetry of the system,
angular momentum $l$ could also be written as $l = 4 + i \mod 6$, as
is shown in the energy spectrum. By inserting a central flux into
this system, we are able to see that a series number of low energy
states, which are 18 states in the case of kagom\'{e} lattice with
$N_s = 42$ sites (corresponding to the system shown in Fig. 4 of the
main text, where we have studied the flux insertion), evolve into each
other without mixing with other higher energy states; see Fig.~\ref{1b_flux}.
When the flux changes its value by $2\pi$, we can clearly see that a
single-particle state with angular momentum $l$ evolves into the state
with angular momentum $l+1 \mod 6$. We will see in the subsequent part
how to infer the flux periodicity of many-body states which are built
from single-particle orbitals with particular flux periodicity.

When a few number of hardcore bosons are loaded into the finite disk
system, they occupy these low-energy states with the restriction that
no more than one particle occupy two consecutive orbitals, which is
stated in the generalized Pauli principle for the $1/2$ FCI/FQAH phases
(Refs.~[8,42] of the main text). Based on this simple picture,
both questions mentioned at the beginning can be well explained without difficulty.

\section*{Angular momentum of many-body ground state}

Following above, the many-body ground state in a $1/2$ FCI/FQAH phase can be
indexed as the ``root configuration'' $|101010 \ldots n_i \ldots \rangle$
(Refs.~[42] of the main text), where $\sum_i n_i =
N_b $ is the particle number and $n_i = 0, 1$ denotes the boson number
occupying the $i$th lowest-energy state. The total angular momentum
$L_{\rm GS}^{(N_b)}$ can be written as $L_{\rm GS}^{(N_b)} =
5+7+\dots+(2N_b+3) = N_b^2 + 4N_b \mod 6$. So $L_{\rm GS}^{(1)} = 5$,
$L_{\rm GS}^{(2)} = 0$, $L_{\rm GS}^{(3)} = 3$, $L_{\rm GS}^{(4)} = 2$,
$L_{\rm GS}^{(5)} = 3$, $L_{\rm GS}^{(6)} = 0$, which is in exact accordance
with our numerical results. We confirmed this GS angular momentum
analysis applies to the situation of kagom\'{e} lattice
(e.g. the 72-site cases in Fig. 3 of the main text). In addition,
this analysis also applies to the honeycomb-lattice Haldane model
(e.g. the 96-site cases in Fig. 2 of the main text).

\section*{Flux periodicity of many-body states}

We now turn to the situation where a central flux is inserted into the
finite-size disk system. Flux insertion is expected to shift the angular momenta
of the many-body states as well as the single-particle ones.
When the flux changes its value by $2\pi$, each single-particle state
moves to the next angular momentum sector. We should keep in mind that many-body
states are built from single-particle orbitals with this particular flux periodicity.
For a disk system with $N_b$ bosons,
this leads to $E_n(L,\Phi+2q\pi)=E_n(L+qN_b,\Phi)$. Using $N_b/6=p/q$,
we obtain $E_n(L+qN_b,\Phi)=E_n(L+6p,\Phi)$, with $p$ and $q$ as
coprime integers. Considering the $C_6$ rotational symmetry of the system,
we finally arrive at the flux periodicity of many-body states (which is stated
in the main text as an empirical fact) as shown below,
\[E_n(L,\Phi+2q\pi)=E_n(L,\Phi)~.\]

\begin{figure}[!htb]
    \includegraphics[scale=0.5]{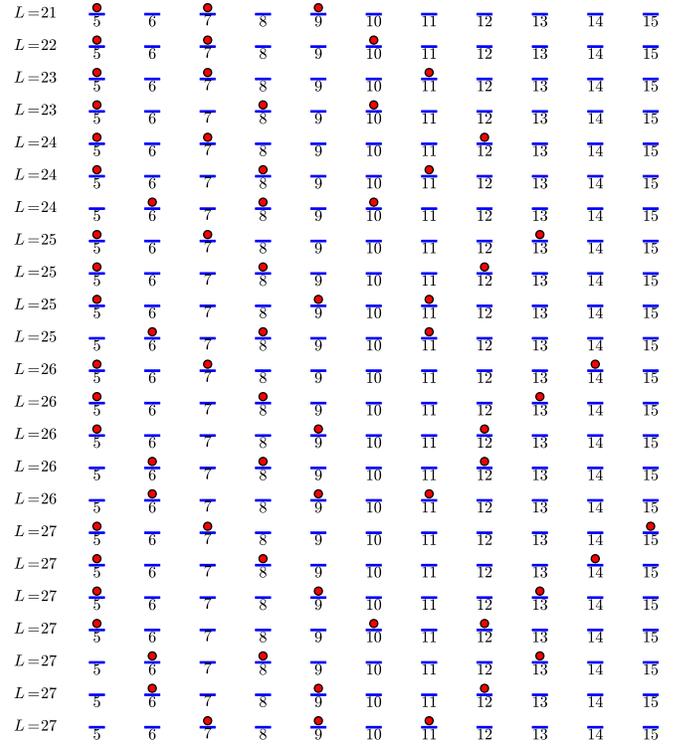}
    \caption{(color online). Illustration of the allowed configurations of $N_b = 3$
    hardcore bosons ruled by the generalized Pauli principle. The short blue lines represent
    the single-particle orbitals with the angular momentum quantum numbers labelled below, and
    the red circles denotes the occupancy of hardcore bosons. The total angular momentum quantum numbers are labelled at the left as ``$L=21$'' etc. }
\label{degeneracy}
\end{figure}

\section*{Degeneracy sequence in low-energy edge excitations}

In the main text of this paper, we numerically find that edge
excitations of finite disk systems in fractional Chern insulators exhibit
a quasi-degeneracy sequence, which turns out to be quite coincident with
that predicted by the chiral Luttinger liquid theory. In this section,
we will show that the degeneracy sequence of these edge excitation
spectra can also be well predicted using the generalized Pauli principle.

The key idea is to count the number of ways to put hardcore bosons
in single-particle orbitals for a certain total angular momentum $L$,
which should be equal to the degeneracy level in sector $L$; the
allowed configurations (Ref.~[42] of the main text) should be ruled by
the generalized Pauli principle. For a finite disk system,
we expect a good match in numerical results when the particle number is small,
where finite-size effect is not so obvious.

Without loss of generality, we discuss a disk system with $N_b = 3$
hardcore bosons. We mentioned in early sections that the root configuration
of ground state should be $|10101000 \ldots \rangle$, where $\ldots$ denotes
subsequent 0's in the case of $N_b = 3$ bosons, and the total angular
momentum should be $L = 5 + 7 + 9 = 21$ (or 3 after $L \mod 6$). Since
this is the only configuration satisfying $L=21$, degeneracy level
$d_{L}$ in this sector should be $d_{L=21}=1$. In the $L=22$ sector, the
only allowed configuration is $|10100100 \ldots \rangle$, where $L=5+7+10 =
22$, so $d_{L=22}=1$. In the $L=23$ sector, the allowed configurations are
$|101000100 \ldots \rangle$, where $L=5+7+11 = 23$, and $|10010100 \ldots \rangle$,
where $L=5+8+10 = 23$, so $d_{L=23}=2$. In this manner, we can deduce
the degeneracy level at each angular momentum for a given number of
particles. An illustration is shown in Fig.~\ref{degeneracy}, which is
$1,1,2,3,4,5,7, ...$ for systems with $N_b=3$, exactly coincident with
what we have obtained from numerical calculations!

Following this spirit, we are able to reproduce the degeneracy sequences for
different numbers of bosons, which is coincident with those predicted by
the chiral Luttinger liquid theory.

To conclude, we are able to heuristically interpret some important features of edge
excitations even in a rather tiny system based on the generalized
Pauli principle. It is also interesting to notice that this principle
applies generally to different lattice models and to different system
sizes, for small boson numbers that the finite-size effect is ignorable.

\end{document}